\documentclass[conference]{IEEEtran}
\IEEEoverridecommandlockouts
\usepackage{amsmath,amssymb,amsfonts}
\usepackage{graphicx}
\usepackage{algorithm}
\usepackage{booktabs}
\usepackage{multirow}
\usepackage{multicol}
\usepackage{algpseudocode}
\usepackage{epstopdf}

\makeatletter       
\renewcommand{\maketag@@@}[1]{\hbox{\m@th\normalsize\normalfont#1}}
\makeatother

\include{amsmath}  

\usepackage{color}

\usepackage{subfigure}

\def\BibTeX{{\rm B\kern-.05em{\sc i\kern-.025em b}\kern-.08em
    T\kern-.1667em\lower.7ex\hbox{E}\kern-.125emX}}
\begin{document}

\title{Non-Binary Polar Coded System for the Two-User Multiple-Access Channel}

{\author{\IEEEauthorblockN{Guan-Chen Liu\IEEEauthorrefmark{1}, Qi-Yue Yu\IEEEauthorrefmark{1}}
\IEEEauthorblockA{\IEEEauthorrefmark{1}Communications Research Center, Harbin Institute of Technology, Harbin 150001, China}

Email: guanchen\_liu@hit.edu.cn, yuqiyue@hit.edu.cn

\thanks{The work presented in this paper was supported in part by the National Natural Science Foundation of China under Grand No. 62071148, and partly supported by the Natural Science Foundation of Heilongjiang Province of China (No. YQ2019F009).}
}}

\maketitle

\begin{abstract}
This paper presents non-binary polar codes for the two-user multiple-access channel (MAC). The bit error rate (BER) performances of the non-binary polar codes with different kernel factors have been investigated in detail to select a proper parameter from ${\rm GF}\left({q}\right)$ for the generator matrix. Furthermore, the successive cancellation decoding for the non-binary polar codes in the two-user MAC is introduced in detail. Simulation results show that the choice of the kernel factors has a significant impact on the block error rate (BLER) performance; moreover, the non-binary polar codes provide a better BLER performance than their binary counterpart in the two-user MAC.
\end{abstract}

\vspace{0.3em}


\section{Introduction}

Polar codes have attracted widespread attention since they can achieve the Shannon limit \cite{Channel polarization}, and much research has been done for polar codes in many aspects. Since 2012, polar codes have been considered for the multiple-access channel (MAC) \cite{polar codes 2 user}-\cite{successive MAC}. Authors in \cite{polar codes 2 user} present a joint polarization for the two-user MAC, which results in five possible transmission models that achieve the dominant face of the capacity region. \cite{polar codes m user} extends the two-user case to the $m$-user case, $m \ge 2$, and deduces the extremal points of the reachable rate region. However, the proposed joint polarization can only reach some of the capacity region instead of all. Paper \cite{compound polar code MAC} proposes compound polar codes for the two-user MAC, which can achieve the whole uniform rate region by changing the decoding order of the joint successive cancellation decoder. In \cite{successive MAC}, the authors utilize the generalized chain rule to construct polar codes for two-user MACs, achieving all the capacity region.

Moreover, the non-binary (NB) polar codes are also an appealing research field because of their flexible structure \cite{polarization for arbitrary}-\cite{Construction two two}. In \cite{polarization for arbitrary}, the authors exploit the randomized construction to polarize the arbitrary input discrete memoryless channel with the binary kernel. In \cite{channel polarization q ary}, the authors present a non-binary kernel form that can be polarized if the input size is the power of a prime and the kernel's parameter is the primitive element of the Galois Field ${\rm GF}\left({q}\right)$. A polarized mapping scheme is discussed in \cite{polar discrete}, which is suited to both the sources and channels. It has been pointed out that multilevel polarization arises when the input size is the power of two \cite{polar q ary}. Then non-binary kernels are considered in \cite{Construction two two} to construct a system for polarized transmission, resulting in a better BLER performance than binary polar codes.

Some literature works have been done on non-binary polar codes for the MAC due to their appealing features. \cite{polar arbitrary DMCs MACs} constructs polar codes using the group structure when the MAC input is arbitrary, achieving the symmetric sum capacity except for some points. In \cite{polar CQ}, the authors present polarization theorems for arbitrary classical-quantum channels with Arikan style transformation, which can be used to construct polar codes for arbitrary classical-quantum MACs with relatively low complexity of encoding and decoding. A channel upgradation polar construction is generalized to the non-binary input MAC case to choose the polarization channel for data transmission \cite{upgradation one}\cite{upgradation two}.

The prior works mainly focus on the binary codes and the achievability of the rate region in theory. Based on the generalized chain rule proposed by \cite{successive MAC}, this paper presents a non-binary polar coded system for the two-user MAC, which mainly exploits the flexible design of the non-binary kernel and the successive cancellation (SC) decoding implementation.

The structure of this paper is arranged as follows. Section II presents the system model of the proposed scheme. A successive cancellation decoding algorithm of non-binary polar codes in the two-user MAC is presented in Section III. In Section IV, the selection of the kernel factors is discussed in detail. The simulation results of the proposed system are shown in Section V, followed by concluding remarks in Section VI.

\section{SYSTEM MODEL}

\begin{figure*}[htbp]
\vspace{-2em}  
\centerline{\includegraphics[width=0.9\textwidth]{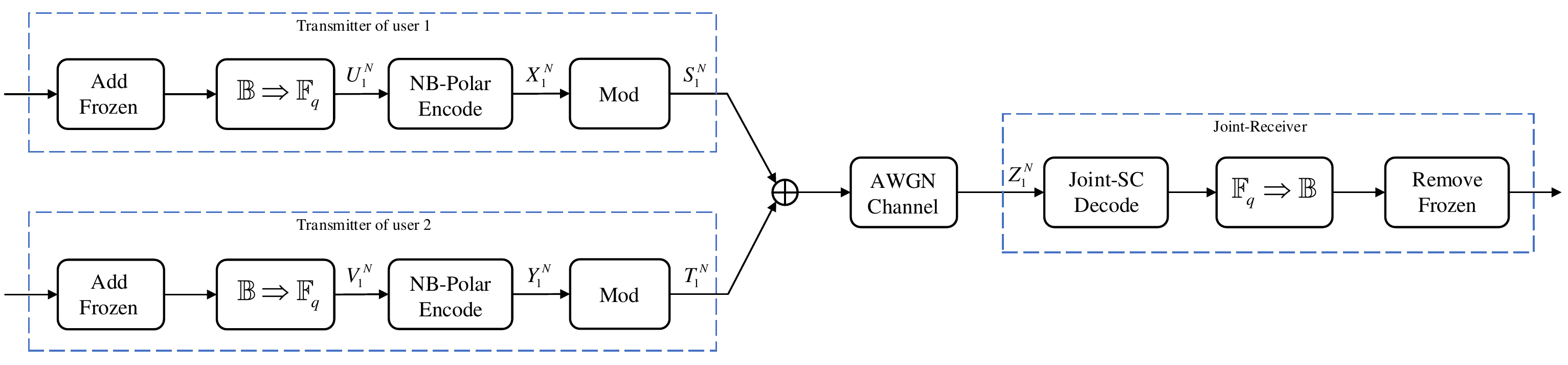}}
\caption{A diagram of the non-binary polar transmission system in the two-user MAC.}
\label{Polar System Model}
\vspace{-1em}  
\end{figure*}

We define $\mathbb{B}$, $\mathbb{N}$ and $\mathbb{R}$ as the binary, natural, and real field, respectively. The Galois field ${\rm GF}\left({q}\right)$ is denoted by ${\mathbb F}_q$, where $q = 2^r$ and $r \in \mathbb{N}$. Considering the element set in $\mathbb{F}_q$ as $ \{ 0 = \gamma^{-\infty},$ $1=\gamma^0,$ $ \gamma, \gamma^2, ..., \gamma^{q-2}\}$, where $\gamma$ is the primitive element of $\mathbb{F}_q$. We follow the notations defined in [1], denoting random variables and the corresponding samples by upper and lower case letters, respectively. Besides, $A_1^N$ stands for a vector $\left(A_1,A_2,..., A_i, ..., A_N \right)$, and $A_i^j$ denotes the subvector $\left(A_i,...,A_j \right)$ for $1 \le i \le j \le N$. Let $\mathcal{N}\left({\mu, \sigma^2}\right)$ represent the Gaussian distribution with the mean $\mu$ and the variance $\sigma^2$.

The system model is shown in Fig. \ref{Polar System Model}. At the transmitter, the frozen bits are added to each user's uncoded information while maintaining the sum rate $R$ for the whole system. Then every $r$ bits are converted to a non-binary symbol of $\mathbb{F}_q$ serially. Denote $N = {2^n}, n \in {\mathbb N}$ as the code length of each block in the non-binary symbol form. The non-binary symbol blocks of users 1 and 2 are respectively defined by $U_1^{N}$ and $V_1^{N}$, where $U_i =\left(U_{i,1}, U_{i,2}, ...,U_{i, t},..., U_{i,r} \right)$, $V_i =\left(V_{i,1}, V_{i,2}, ...,V_{i, t},..., V_{i,r} \right)$, and $U_{i, t}, V_{i,t} \in {\mathbb B}$, $1 \le i \le N$, $1 \le t \le r$. Define ${F}$ as the kernel matrix, given by
\begin{equation} \label{Polar Encode}
{{F} = \left[ {\begin{array}{*{20}{c}}
1&0\\
\alpha &1
\end{array}} \right]~,~ \alpha \in \mathbb{F}_q}.
\end{equation}

Denote $\alpha_u$ and $\alpha_v$ as the kernel factor for users 1 and 2, respectively. The generator matrix of polar codes is defined by ${ G_N} = {B_N}{F^{ \otimes n}}$, where ${F^{ \otimes n}}$ denote the $n$-th Kronecker power of ${F}$ and $B_N$ is the reverse-shuffle. Note that the operations of addition and multiplication are both based on $\mathbb{F}_q$.

Let ${{X}_{1}^{N}}$ and ${{Y}_{1}^{N}}$ denote the encoded polar codewords of users 1 and 2, given by
\begin{equation} \label{Polar Encode}
{X_1^N = U_1^N{G_N}~,~Y_1^N = V_1^N{G_N}},
\end{equation}
with $X_i =\left(X_{i,1}, X_{i,2}, ...,X_{i, t},..., X_{i,r} \right)$, $Y_i =(Y_{i,1}, Y_{i,2}, ...,$ $Y_{i, t},..., Y_{i,r} )$, and $X_{i, t}, Y_{i,t} \in {\mathbb B}$. The encoded polar codewords are then converted to bits. By using BPSK modulation, the modulated signals are $S_1^{N}$ and $T_1^{N}$ for users 1 and 2, respectively. Two users' modulated signals are then transmitted to the multiple-access channel, and the received signals $Z_1^N$ is given by
\begin{equation}
Z_1^N = S_1^N + T_1^N + K_1^N,
\end{equation}
where $Z_i =\left(Z_{i,1}, Z_{i,2}, ...,Z_{i, t},..., Z_{i,r} \right)$, $K_i =(K_{i,1}, K_{i,2},$ $..., K_{i, t},..., K_{i,r} )$, and $Z_{i, t}, K_{i,t} \in {\mathbb R}$. The noise component $K_{i,t}$ satisfies ${\mathcal{N}}\left(0, {{{N_0}} \mathord{\left/
 {\vphantom {{{N_0}} 2}} \right.
 \kern-\nulldelimiterspace} 2}  \right)$.

At the receiver, the successive cancellation decoding is used to recover $U_1^N$ and $V_1^N$, denoted by ${\hat U}_1^N$ and ${\hat V}_1^N$. After converting ${\hat U}_1^N$ and ${\hat V}_1^N$ to bits and removing the frozen bits, the information bits of two users are obtained. By now, the description of the system framework is accomplished. Some definitions for the channel analysis of this system are given next.

Let $W:\mathcal{X} \times \mathcal{Y} \rightarrow \mathcal{Z} $ be a two-user multiple-access channel. Symbol pair $\left( {X_i}, {Y_i} \right)$ can be viewed as the input of $W$, and $W\left( {z_i|x_i,y_i} \right)$ denotes the corresponding transition probability. Set ${W}^{N}$ as $N$ independent uses of ${W}$. Let ${\mathcal{X}^{N}}$ and ${\mathcal{Y}^{N}}$ denote the input of ${W}^{N}$ from user 1 and user 2, while ${\mathcal{Z}^{N}}$ is the output of ${W}^{N}$, then ${W^N}\left( {{z^N_1}|{x^N_1},{y^N_1}} \right){\rm{ = }}\prod\nolimits_{i = 1}^N {W\left( {z_i|x_i,y_i} \right)} $.

Like the case in \cite{successive MAC}, define the combined channel ${{W}_{N}}$ by
\begin{equation} \label{Combined MAC}
{W_N}\left( {{z^N_1}|{u^N_1},{v^N_1}} \right){{ = }}{W^N}\left( {{z^N_1}|{u^N_1}{G_N},{v^N_1}{G_N}} \right).
\end{equation}

Since there is no mutual information loss during the polar transform, thus
\begin{equation} \label{Information chain rule}
I\left( {{Z^N_1};{U^N_1}\!,\!{V^N_1}} \right) \!= \! I\left( {{Z^N_1};{X^N_1}\!,\!{Y^N_1}} \right)\! =\! N \!\cdot\! I\left( {Z;X,Y} \right).
\end{equation}

To construct a polarization channel, $(\ref{Information chain rule})$ is expanded  as
\vspace{-0.2em}    
\begin{equation} \label{Generalized chain rule}
I\left( {{Z^N_1};{U^N_1},{V^N_1}} \right) = \sum\limits_{k = 1}^{2N} {I\left( {{Z^N_1};{E_k}|{E^{k - 1}_1}} \right)},
\vspace{-0.2em}    
\end{equation}
where ${{E}^{2N}_{1}}$ is the permutation of ${{U}^{N}_{1}{V}^{N}_{1}}$ that preserves the monotone order of both ${{U}^{N}_{1}}$ and ${{V}^{N}_{1}}$. Let ${{b}^{2N}_{1}}$ indicate the relative order of ${{E}^{2N}_{1}}$. When ${{E}_{i} \in {U}^{N}_{1}}$, ${{b}_{i} = 0}$; otherwise, ${{b}_{i} = 1}$.

The coordinate channels $W_N^{\left( {b_k,i,j} \right)}$ that correspond to (\ref{Generalized chain rule}) are defined as
\begin{equation}      \label{different b coordinate channel}
\begin{aligned}
W_N^{\left( {{b_k},i,j} \right)}&\left( {z_1^N,e_1^{k - 1}|{e_k}} \right) = \\
&\left\{ {\begin{array}{*{20}{c}}
{W_N^{\left( {0,i,j} \right)}\left( {z_1^N,u_1^{i - 1},v_1^j|{u_i}} \right)},&{{b_k} = 0}\\
{W_N^{\left( {1,i,j} \right)}\left( {z_1^N,u_1^i,v_1^{j - 1}|{v_j}} \right)},&{{b_k} = 1}
\end{array}} \right., 
\end{aligned}
\end{equation}
where $i$ and $j$ stand for the $i$-th symbol of user 1 and the $j$-th symbol of user 2, $0 \le i, j \le N$, $1 \le k=i+j \le 2N$. The coordinate channel transition probability is used in the decoding process as described in the next section.

\section{A JOINT SC DECODING ALGORITHM}

In this section, a joint SC decoding is presented for the NB-polar codes in the two-user MAC. Since the generator matrix $G_N$ is determined by the row vector in ${F^{ \otimes n}}$, thus ${F_N}={F^{ \otimes n}}$ is used to describe the decoding process, ignoring operation $B_N$ that has no impact on the performance.

To calculate $W_N^{\left( {b_k,i,j} \right)}$, we define the split MAC channel $W_N^{\left( {i,j} \right)}$ by
\vspace{-0.6em}    
\begin{small}
\begin{equation} \label{recursive channel}
W_N^{\left( {i,j} \right)}{\!\!}\left( \!{{z^N_1}\!,{u^{i - 1}_1}\!\!,{v^{j - 1}_1}|{u_i},{v_j}}\! \right) {\!\!}={\!\!\!\!\!\!} {\sum\limits_{u_{i + 1}^N,v_{j + 1}^N}{\! \!\! \!}{\left({\!} {\frac{1}{{{q^{N {\!}-{\!} 1}}}}} {\!}\right)} ^2}{\!\!\!}{W_N}{\!\!}\left( \!{{z^N_1}|{u^N_1}\!,\!{v^N_1}}\! \right){\!\!}.
\vspace{-0.9em}    
\end{equation} 
\end{small}

When considering the decoding order, the single user transition probability is derived as
\vspace{-0.3em}    
\begin{equation}      \label{single decoding u}
\begin{aligned}
W_N^{\left( {0,i,j} \right)}&\left( {z_1^N,\hat u_1^{i - 1},\hat v_1^j|{u_i}} \right) = \\
&\left\{ {\begin{array}{*{20}{c}}
{{\!}{\!}{\!}{\!}{\!}\sum\limits_{v1} {\frac{1}{q}W_N^{\left( {i,1} \right)}\left( {z_1^N,\hat u_1^{i - 1}|{u_i},{v_1}} \right)} }&{{\rm{if}}\;j = 0}\\
{{\!}{\!}{\!}{\!}\frac{1}{q}W_N^{\left( {i,j} \right)}\left( {z_1^N,\hat u_1^{i - 1},\hat v_1^{j - 1}|{u_i},{\hat v_j}} \right)}&{{\rm{if}}\;j > 0}
\end{array}} \right.{\!}{\!},
\end{aligned}
\vspace{-0.2em}    
\end{equation}
\begin{equation}      \label{single decoding v}
\begin{aligned}
W_N^{\left( {1,i,j} \right)}&\left( {z_1^N,\hat u_1^i,\hat v_1^{j - 1}|{v_j}} \right) = \\
&\left\{ {\begin{array}{*{20}{c}}
{{\!}{\!}{\!}{\!}{\!}\sum\limits_{u1} {\frac{1}{q}W_N^{\left( {1,j} \right)}\left( {z_1^N,\hat v_1^{j - 1}|{u_1},{v_j}} \right)} }&{\!}{{\rm{if}}\;i = 0}\\
{{\!}{\!}{\!}{\!}\frac{1}{q}W_N^{\left( {i,j} \right)}\left( {z_1^N,\hat u_1^{i - 1},\hat v_1^{j - 1}|{\hat u_i},{v_j}} \right)}&{\!}{{\rm{if}}\;i > 0}
\end{array}} \right.{\!}{\!}.
\end{aligned}
\vspace{-0.5em}    
\end{equation}

\begin{figure}[t]
\centering
\includegraphics[width = 0.495\textwidth]{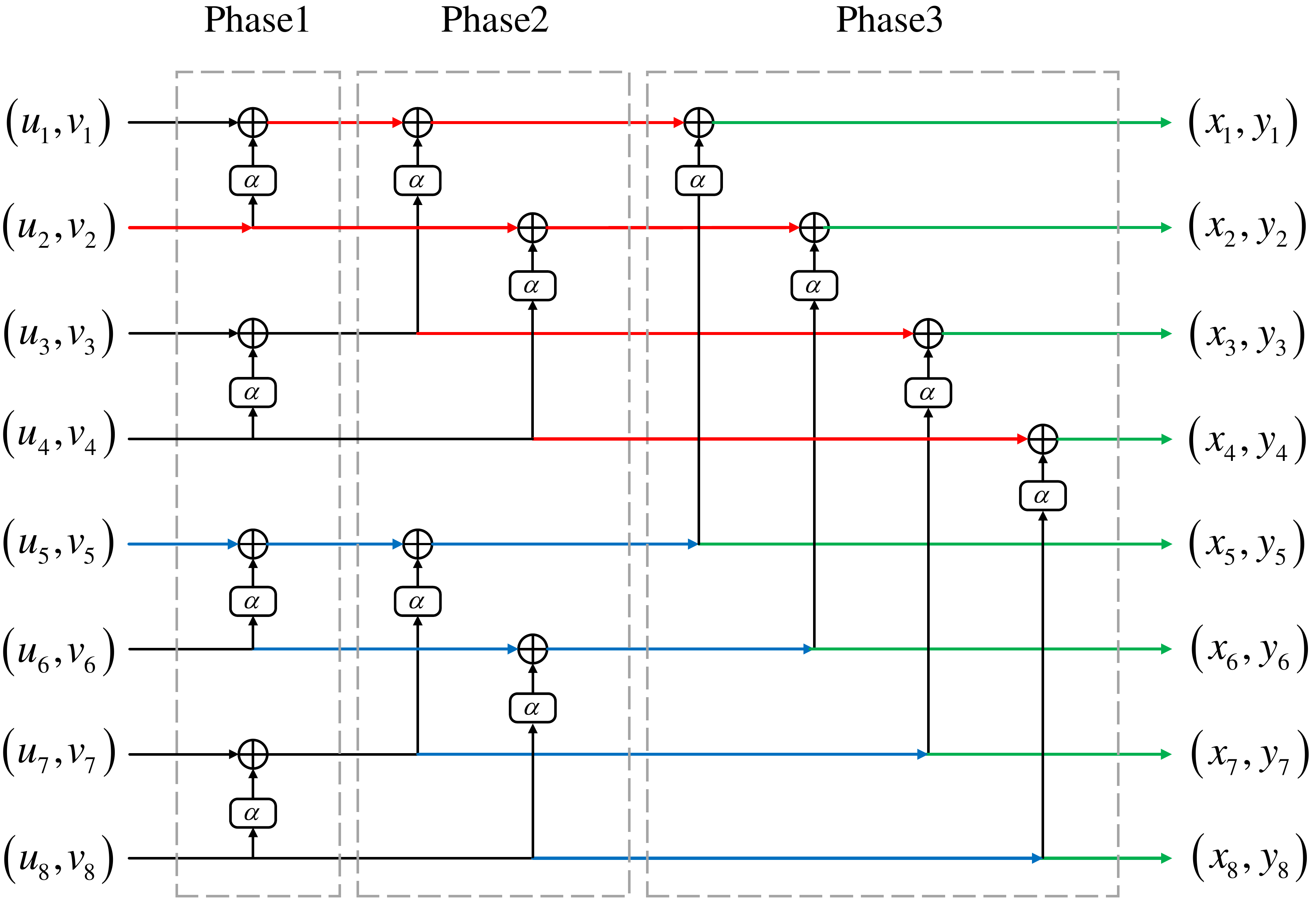} 
\caption{An example of the decoding path with three phases, where $N=8$.}
\vspace{-1.5em}
\label{Decoding_path}
\end{figure}
\vspace{0em}

The transition probability of the split MAC channel $W_N^{\left( {i,j} \right)}$ can be calculated recursively due to the recursive structure. For simplicity, here we use the notations similar to \cite{successive MAC}. Let $\dot u_i \buildrel \Delta \over = {u_{2i - 1}} + \alpha_u \cdot {u_{2i}}$ and $\ddot u_i  \buildrel \Delta \over = {u_{2i}}$, and define $\dot u_1^i = u_{1,o}^{2i} + {\alpha _u} \cdot u_{1,e}^{2i}$ and $\ddot u_1^i = u_{1,e}^{2i}$, where the subscripts $o$ and $e$ denote the subvector with odd and even indices that are similar to $\dot{v}_1^i$ and $\ddot{v}_1^i$, respectively. Define
\begin{equation}      \label{abbreviation CD}
\begin{aligned}
D_N^{\left( {i,j} \right)}{\!}\left({\!} {z_1^{2N}{\!}{\!},u_1^{2i},v_1^{2j}} \right) {\!}{\!}\buildrel \Delta \over ={\!}{\!}~ &W_N^{\left( {i,j} \right)}{\!}\left( {\!}{z_1^N{\!},\dot u_1^{i - 1}{\!}{\!},\dot v_1^{j - 1}|{{\dot u}_i},{{\dot v}_j}} {\!}\right)\\
\cdot~ &W_N^{\left( {i,j} \right)}{\!}\left( {\!}{z_{N + 1}^{2N},\ddot u_1^{i - 1}{\!}{\!},\ddot v_1^{j - 1}|{{\ddot u}_i},{{\ddot v}_j}} {\!}\right){\!}.
\end{aligned}
\end{equation}

Then the recursive equations are given by
\begin{align}  
W_{2N}^{\left( {2i - 1,2j - 1} \right)}&\left( {z_1^{2N},u_1^{2i - 2},v_1^{2j - 2}|{u_{2i - 1}},{v_{2j - 1}}} \right) \notag \\
 = \sum\limits_{{u_{2i}},{v_{2j}}} {\frac{1}{{{q^2}}}} D_N^{\left( {i,j} \right)}&\left( {z_1^{2N},u_1^{2i},v_1^{2j}} \right), \label{decoding E1} \\
W_{2N}^{\left( {2i,2j - 1} \right)}&\left( {z_1^{2N},u_1^{2i - 1},v_1^{2j - 2}|{u_{2i}},{v_{2j - 1}}} \right) \notag \\
 = \sum\limits_{{v_{2j}}} {\frac{1}{{{q^2}}}} D_N^{\left( {i,j} \right)}&\left( {z_1^{2N},u_1^{2i},v_1^{2j}} \right), \label{decoding E2} \\
W_{2N}^{\left( {2i - 1,2j} \right)}&\left( {z_1^{2N},u_1^{2i - 2},v_1^{2j - 1}|{u_{2i - 1}},{v_{2j}}} \right)  \notag \\
 = \sum\limits_{{u_{2i}}} {\frac{1}{{{q^2}}}} D_N^{\left( {i,j} \right)}&\left( {z_1^{2N},u_1^{2i},v_1^{2j}} \right), \label{decoding E3} \\
W_{2N}^{\left( {2i,2j} \right)}&\left( {z_1^{2N},u_1^{2i - 1},v_1^{2j - 1}|{u_{2i}},{v_{2j}}} \right) \notag  \\
 = \frac{1}{{{q^2}}}D_N^{\left( {i,j} \right)}&\left( {z_1^{2N},u_1^{2i},v_1^{2j}} \right).\label{decoding E4}
\end{align}

According to (\ref{decoding E1})-(\ref{decoding E4}), the recursive transform implies a decoding path for each split channel. For example, the decoding path of $W_8^{\left( {2,5} \right)}\left( {{z^8_1},{{\hat u}_1},{{\hat v}^4_1}|{u_2},{v_5}} \right)$ is shown in Fig. \ref{Decoding_path}, where red lines, blue lines, and green lines represent the decoding path for user 1, user 2, and both two users, respectively. Assuming that the phase index $p$ increases from the input side to the output side in the polar encoder, $1\le p\le n$. In each phase, two users' decoding paths are combined to calculate the corresponding probabilities.

When we decode symbol $U_i \in {U}_{1}^{N}$, $1 \le i \le N$, all the known decisions $\hat V$ are viewed as auxiliary symbols, similarly to the symbol $V_i \in {V}_{1}^{N}$. Since the decoding process of ${U}$ and ${V}$ are reciprocity, we use the decoding process of ${U}$ with the auxiliary information ${V}$ for the analysis in the following.

\begin{figure}[t]   
\centering
\vspace{1em}   
\includegraphics[width = 0.25\textwidth]{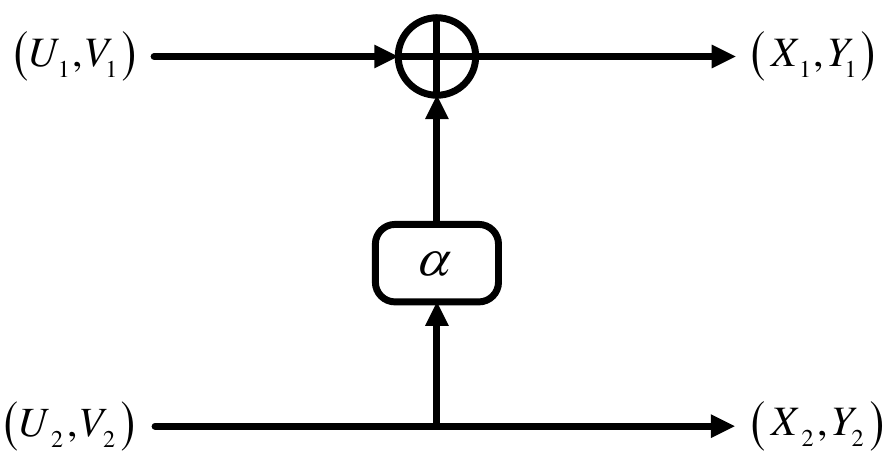}
\caption{The basic kernel (or combined) structure of the NB-polar generator.}
\vspace{-1em}   
\label{Combine Basic Structure}
\end{figure}

\begin{figure}[t]
\centering
  \subfigure[$f_{0,v}$]{
    \label{u f 0v} 
    \includegraphics[width=0.23\textwidth]{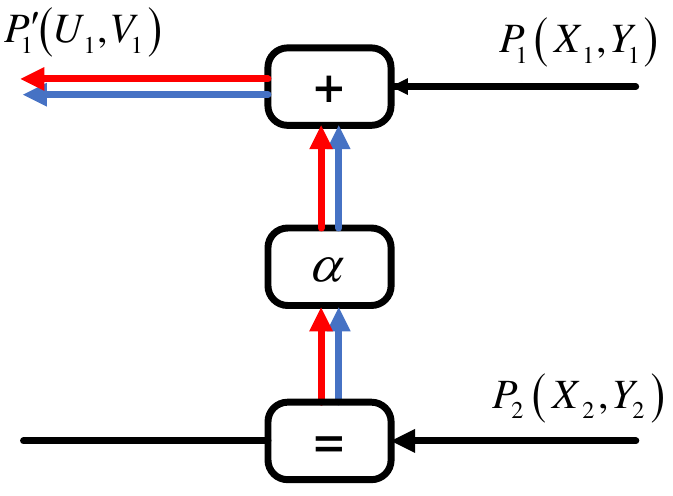}}
  \subfigure[$g_{0,v}$]{
    \label{u g 0v} 
    \includegraphics[width=0.23\textwidth]{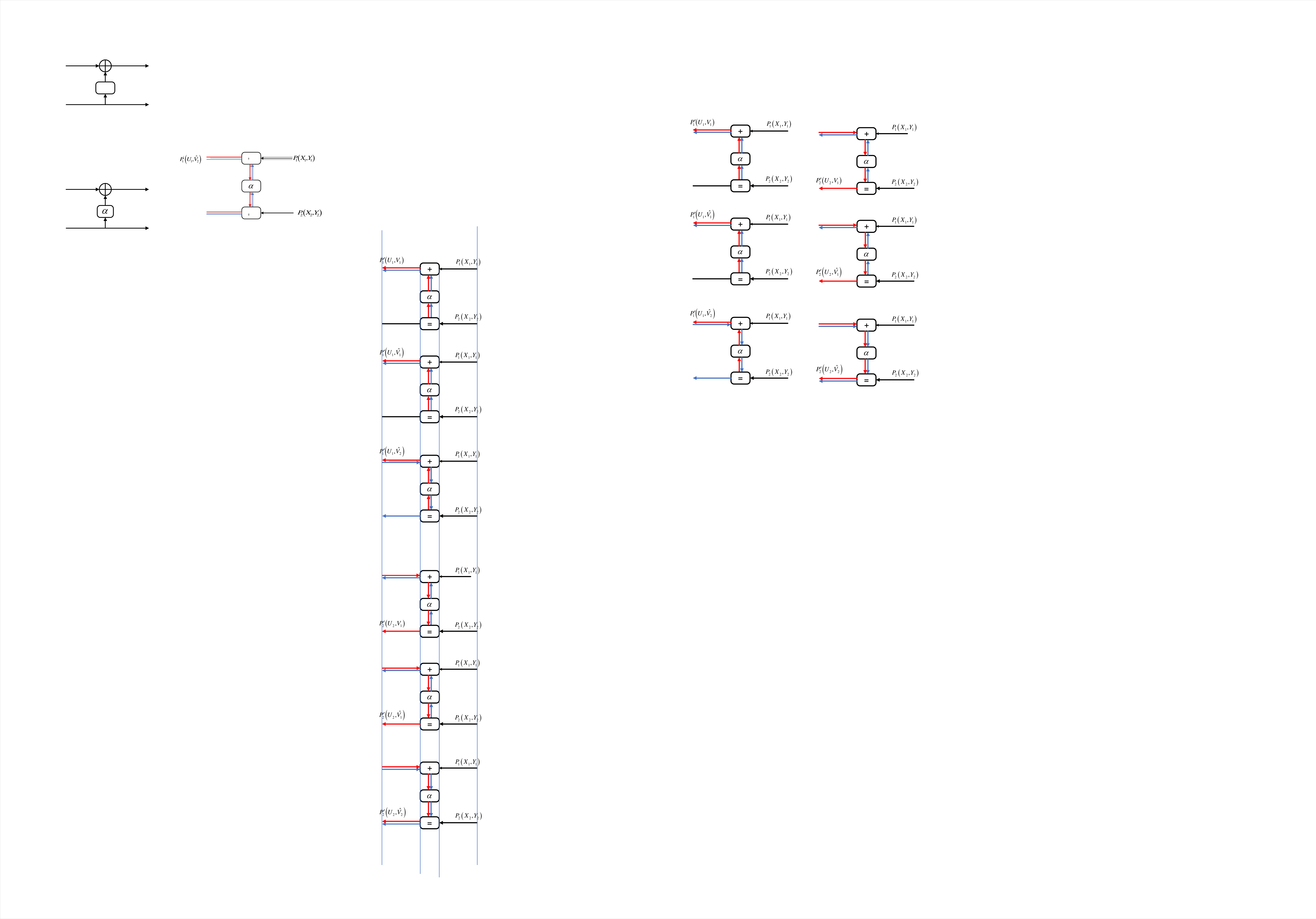}}
  \subfigure[$f_{1,v}$]{
    \label{u f 1v} 
    \includegraphics[width=0.23\textwidth]{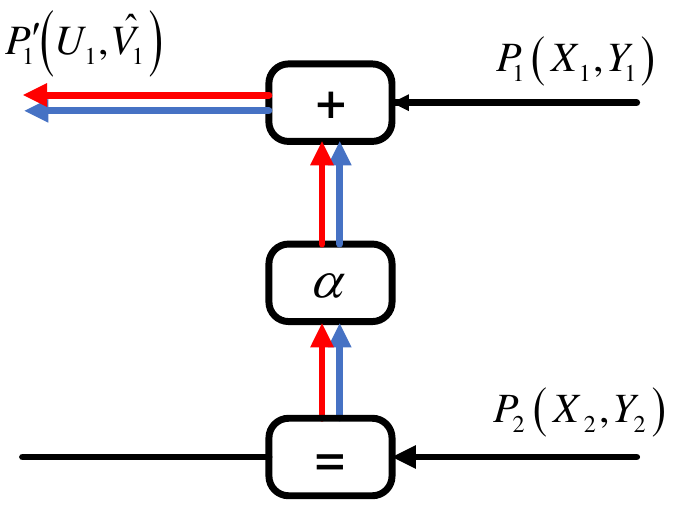}}
  \subfigure[$g_{1,v}$]{
    \label{u g 1v} 
    \includegraphics[width=0.23\textwidth]{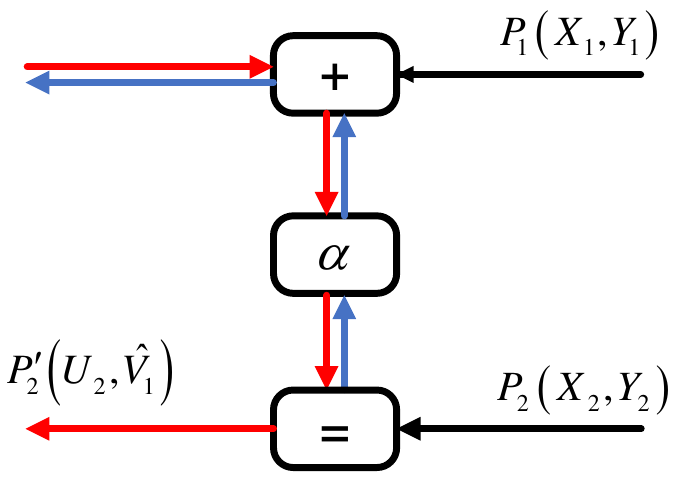}}
  \subfigure[$f_{2,v}$]{
    \label{u f 2v} 
    \includegraphics[width=0.23\textwidth]{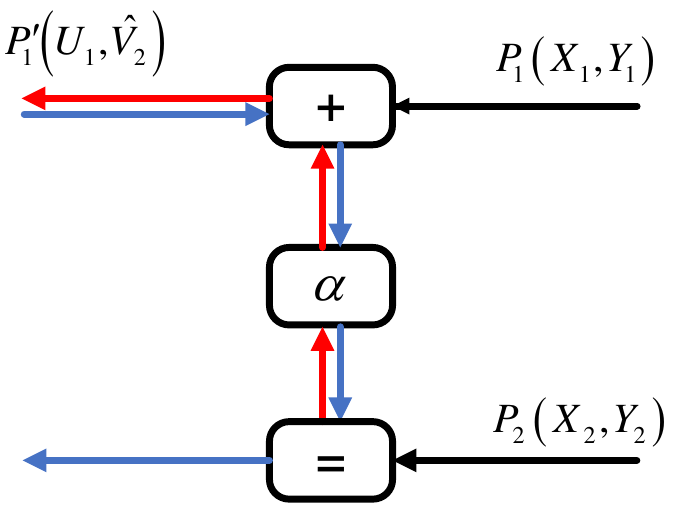}}
  \subfigure[$g_{2,v}$]{
    \label{u g 2v} 
    \includegraphics[width=0.23\textwidth]{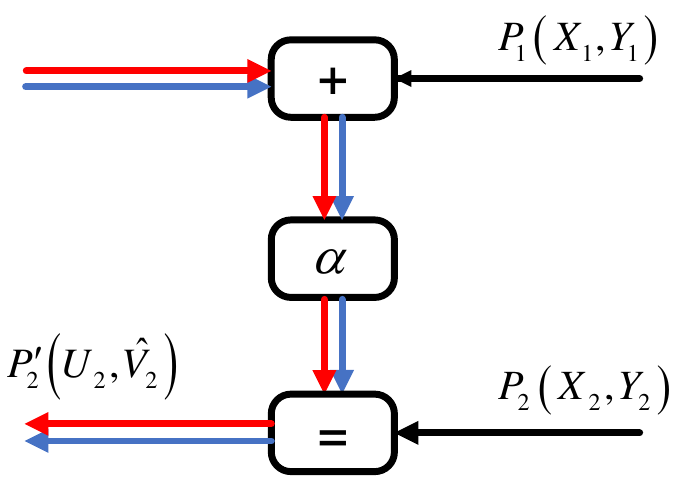}}
  \caption{Six different $f$ and $g$ calculations.}\vspace{-0.5em}
  \label{f g all} 
\end{figure}

The transition probability transform corresponding to the decoding path of each phase is based on the combined basic structure, as shown in Fig. \ref{Combine Basic Structure}. The calculation of probability is divided into two categories: $f$ calculation and $g$ calculation, which correspond to decode $U_1$ and $U_2$ of the combined basic structure, respectively. According to (\ref{decoding E1})-(\ref{decoding E4}), considering the number of auxiliary symbols, there are six types of calculations, denoted by $f_{0, v}\left({P_1, P_2}\right)$, $f_{1, v}\left({P_1, P_2, V_1}\right)$, $f_{2, v}\left({P_1, P_2, V_1, V_2}\right)$ and $g_{0, v}\left({P_1, P_2, U_1}\right)$, $g_{1, v}\left({P_1, P_2, U_1, V_1}\right)$, $g_{2, v}\left({P_1, P_2, U_1, V_1, V_2}\right)$, respectively, as shown in Fig. {\ref {f g all}}. The subscript represents the number of the available auxiliary symbols of $V_1^2$. Based on the combined basic structure and calculations, the decoding process of each phase is similar to the single-user case.

A complete binary tree $\mathbb{T}$ of depth $n+1$ is defined first to indicate the SC decoding process \cite{SC Tree}. Given a node $m$, define its location of the decoding tree by the vector ${\left({d_m, c_m}\right)}$, representing the $c_m$-th node of the $d_m$-th depth, $1 \le d_m \le n+1, 1 \le c_m \le 2^{{d_m}-1}$. Let the node's parent node, left and right child node be $p_m$, $m_l$, and $m_r$, respectively.

There are four kinds of information stored in the ${\left({d_m, c_m}\right)}$-th node $m$: the probability matrices $\Phi_m$, decision symbol vector $\beta_m$, auxiliary symbol vector ${\theta_m}$, and child nodes set $\mathbb{C}_m$. Let $\Phi_m \left[{l}\right]$ be the $l$-th row in $\Phi_m$, and $\beta_m \left[{l}\right], {\theta_m}\left[{l}\right]$ be the $l$-th element in $\beta_m, \theta_m$, where $1 \le l \le \rho_m$, $\rho_m = 2^{\left({n - d_m + 1}\right)}$.

For a two-user decoding algorithm, two decoding trees are defined as $\mathbb{T}_u$ and $\mathbb{T}_v$, respectively. As stated in \cite{successive MAC}, for simplicity, ${b}_{2N}=\left({{0}^{M}{1}^{N}{0}^{N - M}}\right)$ is defined as our decoding order, where $1 \le M \le N$. The decoding order can be naturally divided into three stages: decode ${U}_{1}^{M}$ in stage {\uppercase\expandafter{\romannumeral1}}, decode ${V}_{1}^{N}$ in stage {\uppercase\expandafter{\romannumeral2}}, and decode ${U}_{M+1}^{N}$ in stage {\uppercase\expandafter{\romannumeral3}}. Next, we will give a detailed description of the decoding process, consisting of six steps.

\emph{Step 1:} Initialize $\Phi_m$ and $\mathbb{C}_m$ for stage {\uppercase\expandafter{\romannumeral1}}.

The transition probability $P\left({z|x, y}\right)$ is initialized to the root node's $\Phi_m$ of $\mathbb{T}_u$, denoted as
\vspace{-0.5em}   
\begin{equation}
\Phi_m \left[{l}\right] {\!}{\!}={\!} \frac{1}{{{{\left( {\sqrt {2\pi } \sigma } \right)}^r}}}{\!}\prod\limits_{t = 1}^r  \exp {\!}{\!} \left( {\!}{\!}{ - \frac{{{{\left( {{z_{l,t}} {\!}+{\!} 2{x_{l,t}} {\!}+{\!} 2{y_{l,t}} {\!}-{\!} 2} \right)}^2}}}{{2{\sigma ^2}}}} {\!}\right){\!},
\end{equation}
where $1 \le l \le N$. $\mathbb{C}_m$ is initialized with a set for the internal node, including $m_l$ and $m_r$. For the leaf node, $\mathbb{C}_m$ is initialized with an empty set $\phi$.

\emph{Step 2:} The decoding process for stage {\uppercase\expandafter{\romannumeral1}}.

Update $\Phi_m, \beta_m, \mathbb{C}_m$ in $\mathbb{T}_u$. The node passing routine and update operation are relayed on the cardinality of $\mathbb{C}_m$, denoted as $\left|{\mathbb{C}_m}\right|$, and there are three cases.

If $\left|{\mathbb{C}_m}\right|=2$, change to its left child node $m_l$ and remove $m_l$ from $\mathbb{C}_m$. Let $1 \le l \le \frac{\rho_m}{2}$ and update $\Phi_{m_l}$ as
\vspace{-0.3em}    
\begin{equation} 
\Phi_{m_l}\left[ l \right] = {f_{0,v}}\left( {\Phi_{m}\left[ l \right],\Phi_{m}\left[ {l + \frac{\rho_m}{2}} \right]} \right).
\end{equation}

If $\left|{\mathbb{C}_m}\right|=1$, change to its right child node $m_r$ and remove $m_r$ from $\mathbb{C}_m$. Update $\Phi_{m_r}$ as
\vspace{-0.3em}      
\begin{equation} 
\Phi_{m_r}\left[ l \right] = {g_{0,v}}\left( {\Phi_{m}\left[ l \right],\Phi_{m}\left[ {l + \frac{\rho_m}{2}} \right]}, \beta_{ml} \left[{l}\right] \right).
\vspace{-0.2em}    
\end{equation}

If $\left|{\mathbb{C}_m}\right|=0$, change to its parent node $p_m$. There are two situations for this case. On one hand, when $m$ is an internal node, let $m_l$ denote the left child node of $p_m$, update $\beta_{p_m}$ as
\begin{equation}
{\beta _{{p_m}}}\left[ l \right]{\rm{ = }}\left\{ {\begin{array}{*{20}{c}}
{{\beta _{{m_l}}}\left[ l \right] + \alpha  \cdot {\beta _m}\left[ l \right]}&{1 \le l \le {\rho _m}}\\
{{\beta _m}\left[ {l - {\rho _m}} \right]}&{{\rho _m} + 1 \le l \le 2{\rho _m}}
\end{array}} \right..
\end{equation}

On the other hand, when $m$ is a leaf node, make a decision on $\Phi_m$ and update $\beta_m$ as
\vspace{-0.4em}    
\begin{equation}
\beta_m = \mathop {\arg \max }\limits_{u \in \mathbb{F}_q} \sum\limits_{v \in \mathbb{F}_q} \Phi_m .
\vspace{-0.5em}    
\end{equation}

This stage is terminated when ${U}_{M}$ has been decoded.

\emph{Step 3:} Initialize $\theta_m$ for stage {\uppercase\expandafter{\romannumeral2}}.

Some of $\beta_m$ in $\mathbb{T}_u$ is initialized to $\theta_m$ in $\mathbb{T}_v$. Let $J=M$. For the leaf node, initialize $\theta_m$ as
\begin{equation}
\theta_m \left({n + 1, J}\right) = \beta_m\left({n + 1, J}\right).
\end{equation}

Then for the non-leaf node, let $0 \le k \le n-1$, $\theta_m$ is updated from the leaf side to the root side recursively

\begin{equation}   
\begin{aligned}
\theta_m \left({n - k , J}\right) = \beta_m\left({n- k , J }\right), J = \left\lceil {\frac{J}{2}} \right\rceil.
\end{aligned}
\end{equation}

\emph{Step 4:} The decoding process for stage {\uppercase\expandafter{\romannumeral2}}.

This process is similar to step 2. The difference exists in update $\Phi_m$ with $\theta_m$.

If $\left|{\mathbb{C}_m}\right|=2$, update $\Phi_{m_l}$ as
\begin{equation}  \label{three f calculation}
{\Phi _{{m_l}}}\left[ l \right] = f\left( {{\Phi _m}\left[ l \right],{\Phi _m}\left[ {l + \frac{\rho_m}{2}} \right],{\theta _{{m_l}}}\left[ l \right],{\theta _{{m_r}}}\left[ l \right]} \right).
\end{equation}

If $\left|{\mathbb{C}_m}\right|=1$, update $\Phi_{m_r}$ as
\begin{equation}   \label{three g calculation}
{\Phi _{{m_r}}}{\!}\left[ l \right] {\!}={\!} g{\!}\left( {\!}{{\Phi _m}{\!}\left[ l \right]{\!},{\Phi _m}\left[ {l {\!}+{\!} \frac{\rho_m}{2}} \right]{\!},{\beta _m}{\!}\left[ l \right],{\theta _{{m_l}}}{\!}\left[ l \right],{\theta _{{m_r}}}{\!}\left[ l \right]} {\!}\right){\!},
\end{equation}
where $1 \le l \le \frac{\rho_m}{2}$. We choose different calculation types to calculate (\ref{three f calculation}) and (\ref{three g calculation}) according to the value of $\theta_{ml}$ and $\theta_{mr}$. This stage is terminated when $V_N$ is decoded successfully.

\emph{Step 5:} Initialize $\theta_m$ for stage {\uppercase\expandafter{\romannumeral3}}.

Some of $\beta_m$ in $\mathbb{T}_v$ is initialized to $\theta_m$ in $\mathbb{T}_u$. Set $J = N$ and execute the recursive update in step 3, then the initialization for $\theta_m$ is accomplished.

\emph{Step 6:} The decoding process for stage {\uppercase\expandafter{\romannumeral3}}.

This process is similar to step 4, and this stage is terminated when $U_N$ is decoded successfully.
 
The algorithm is summarized in Algorithm 1, which yields $O\left({q\cdot N\cdot\log N}\right)$ operations, a slight increase in contrast with the $O(N\cdot\log N)$ complexity of its binary counterpart.

\begin{algorithm}[t]	  
    \caption{A Successive Cancellation decoding algorithm of NB-polar in the two-user MAC.}
    \label{AlgorithmOne}
    \begin{algorithmic}[1]
        \State \textbf{Input}: ${\mathbb{T}_{u}}$, ${\mathbb{T}_{v}}$, $n$, $M$
        \State \textbf{Initialize for stage {\uppercase\expandafter{\romannumeral1}}.}
		 \State Initialize $\Phi_m$, $\mathbb{C}_m$
        \State \textbf{Decoding process for stage {\uppercase\expandafter{\romannumeral1}}}
	        \For{$i = 1 : M$}
					\While{$d_m \neq n+1$}
						\State Update $\Phi_m, \beta_m, \mathbb{C}_m$ in $\mathbb{T}_u$. 
						\State Change to the next node.
					\EndWhile
					\State ${\hat u_i} = \mathop {\arg \max }\limits_{u \in \mathbb{F}_q} \sum\limits_{v \in \mathbb{F}_q} \Phi_m $. 
					\State Update $\beta_m$ and change to the next node.
 			  \EndFor
        \State \textbf{Initialize for stage {\uppercase\expandafter{\romannumeral2}}}
		 \State Initialize $\Phi_m$, $\mathbb{C}_m$, $\theta_m$
        \State \textbf{Decoding process for stage {\uppercase\expandafter{\romannumeral2}}}
	        \For{$i = 1 : N$}
					\While{$d_m \neq n+1$}
						\State Update $\Phi_m, \beta_m, \mathbb{C}_m$ in $\mathbb{T}_u$. 
						\State Change to the next node.
					\EndWhile
					\State ${\hat v_i} = \mathop {\arg \max }\limits_{v \in \mathbb{F}_q} \Phi_m$. 
					\State Update $\beta_m$ and change to the next node.
 			  \EndFor
         \State \textbf{Initialize for stage {\uppercase\expandafter{\romannumeral3}}}
		 \State Initialize $\theta_m$
        \State \textbf{Decoding process for stage {\uppercase\expandafter{\romannumeral3}}}
			\For{$i = M+1 : N$}
					\While{$d_m \neq n+1$}
						\State Update $\Phi_m, \beta_m, \mathbb{C}_m$ in $\mathbb{T}_u$. 
						\State Change to the next node.
					\EndWhile
					\State ${\hat u_i} = \mathop {\arg \max }\limits_{u \in \mathbb{F}_q} \Phi_m$. 
					\State Update $\beta_m$ and change to the next node.
 			  \EndFor
       \State \textbf{Output}: ${{\hat u}_{1}^{N}, {\hat v}_{1}^{N}}$.
   \end{algorithmic}

\end{algorithm}

\section{THEORETICAL ANALYSIS OF THE KERNEL}

This section focuses on the effect of the kernel. Since the reliability is relatively low in stage {\uppercase\expandafter{\romannumeral1}}, the performance is mainly determined by the kernels of stages {\uppercase\expandafter{\romannumeral2}} and {\uppercase\expandafter{\romannumeral3}}. First, we consider $\alpha_u$ in stage {\uppercase\expandafter{\romannumeral3}} since $\alpha_v$ has no effect when all the auxiliary symbols $V_1^N$ are already known. Then we optimize $\alpha_v$ in stage {\uppercase\expandafter{\romannumeral2}}, assuming $\alpha_u$ is fixed.

\subsection{The Kernel of Stage {\uppercase\expandafter{\romannumeral3}}}

The probability transform of stage {\uppercase\expandafter{\romannumeral3}} is equivalent to that of the single-user basic structure, as shown in Fig. \ref{Combine Basic Structure}, where $U_1, U_2$ and $X_1, X_2$ are the input and output, respectively.

Set $u_{1} = 0$, the transition probability can be written as
\vspace{-0.3em}    
\begin{equation}
{P_i}\left( {z_i|x_i} \right) {\!}={\!} \frac{1}{{{{\left( {\sqrt {2\pi } \sigma } \right)}^r}}}{\!}\prod\limits_{t = 1}^r{\!} {\exp {\!}\left( {\!}{ - \frac{{{{\left( {{z_{i, t}} + {2x_{i, t}} - 1} \right)}^2}}}{{2{\sigma ^2}}}} \right)}, 
\end{equation}
where $1 \le i \le 2$, $\sigma^2 = {{{N_0}} \mathord{\left/
 {\vphantom {{{N_0}} 2}} \right.
 \kern-\nulldelimiterspace} 2}$. The coordinate channel transition probability of $u_2$ is
\begin{small}     
\begin{equation}      
\begin{aligned}
&P\left( {z_1^2,{u_1}|{u_2}} \right) = {P_1}\left( {{z_1}|{x_1}} \right){P_2}\left( {{z_2}|{x_2}} \right) \\
 =& \frac{1}{{{{\left( {\sqrt {2\pi } \sigma } \right)}^{2r}}}}\prod\limits_{i = 1}^2 {\prod\limits_{t = 1}^r {\exp \left( { - \frac{{{{\left( {{z_{i,t}} + 2{x_{i,t}} - 1} \right)}^2}}}{{2{\sigma ^2}}}} \right)} } .
\end{aligned}
\end{equation}
\end{small}

Acccording to the maximum likelihood detection, the decoded $\hat{u}_2$ is
\begin{equation}
{\hat u_2} = \mathop {\arg \max }\limits_{{u_2} \in \mathbb{F}_q} P\left( {z_1^2,{u_1}|{u_2}} \right), 
\end{equation}

Define $L_s = P\left( {z_1^2,{u_1}|\gamma^s} \right)$, ${\gamma ^s} \in {\mathbb{F}_q}$. Assuming $\tilde u_2 = \gamma^s$ is transmitted, the probability of a correct decision is given by
\begin{equation} \label{correct decision}
{P_c} = P\left[ {{L_s}~\textgreater~ {L_a},{\rm for~all~}{\gamma^a} \in {\mathbb{F}_q}\backslash \left\{ \gamma ^s \right\}|{u_2} = {\gamma ^s}} \right].
\end{equation}

Assume the detection of $\gamma^s$ is independent, (\ref{correct decision}) becomes 
\begin{equation}    \label{independent pc} 
{P_c} = {\!}{\!}{\!}{\!}\prod\limits_{{\gamma ^a} \in {\mathbb{F}_q},{\gamma ^a} \ne {\gamma ^s}} {\!}{\!}{\!}{P\left[ {{L_s} > {L_a}|{u_2} = {\gamma ^s}} \right].} 
\end{equation}

\vspace{-0.4em}    
Let $\bar u_2$ denote $\gamma^a$, the term in (\ref{independent pc}) can be organized as
\vspace{-0.5em}    
\begin{equation}    \label{generalize nonsense}
\sum\limits_{i = 1}^2 {\sum\limits_{t = 1}^r {\left( {{{\bar x}_{i, t}}^2 {\!}-{\!} {\tilde x_{i, t}}^2 {\!}+{\!} {\bar x_{i, t}}{z_{i, t}} {\!}-{\!} {{\tilde x}_{i, t}}{z_{i, t}}}{\!}+{\!} {\tilde x_{i, t}} {\!}-{\!} {\bar x_{i, t}} \right)    } } {\!} > {\!}0,
\end{equation}
where $\bar x_1, \bar x_2$ and $\tilde x_1, \tilde x_2$ correspond to $\bar u_2$ and $\tilde u_2$, respectively. Since $z_{i, t} \sim {\mathcal{N}}\left({-2{\tilde x_{i, t}}+1, \sigma ^{2}}\right)$, the left side of (\ref{generalize nonsense}) is also a Gaussian variable $Q \sim {\mathcal{N}}\left({\mu_q, \sigma_q^2}\right)$, where 
\begin{equation}
\mu_q = \sum\limits_{i = 1}^2 {\!}{\sum\limits_{t = 1}^r {{{\left( {{{\tilde x}_{i,t}} {\!}-{\!} {{\bar x}_{i,t}}} \right)}^2}} }, \sigma_q^2 = \sum\limits_{i = 1}^2 {\!}{\sum\limits_{t = 1}^r {\left( {\tilde x_{i,t}^2 {\!}+{\!} \bar x_{i,t}^2} \right)} } {\sigma^2}, 
\end{equation}
and the probability is calculated by integrating the probability density function (PDF) of $Q$. The average BER is calculated by going through all $u_2$. Then we choose the kernel factor $\alpha_u$ that can provide the best BER performance.

\subsection{The Kernel of Stage {\uppercase\expandafter{\romannumeral2}}}

Assuming that $\alpha_u$ is fixed, then we select the optimal $\alpha_v$. Set $u_{1}$ $= 0$ and $v_{1}$ $= 0$, the transition probability is
\vspace{-0.3em}    
\begin{small}
\begin{equation} \label{probability hard}
{\!}{\!}{\!}{P_i}{\!}\left( {z_i|x_i,y_i} \right) {\!}{\!}={\!} \frac{1}{{{{\left( {\sqrt {2\pi } \sigma } \right)}^r}}}{\!}\prod\limits_{t = 1}^r{\!} {\exp {\!}{\!}\left( {\!}{\!}{ - \frac{{{{\left( {{z_{i, t}} {\!}+{\!} {2x_{i, t}} {\!}+{\!} {2y_{i, t}}{\!}-{\!}2} \right)}^2}}}{{2{\sigma ^2}}}} {\!}\right)}{\!},
\end{equation}
\end{small}
Define $\check x_1$, $\check x_2$ by $\left\{ {\check x_1, \check x_2 \in \mathbb{F}_q} | \check x_1 + \alpha_u \cdot \check x_2 = u_1   \right\}$, the split channel transition probability of $u_1, v_2$ is given by
\begin{equation}  \label{sum hard probability}
\begin{array}{*{20}{l}}
{P\left( {z_1^2,{v_1}|{u_1},{v_2}} \right) = }\\
{\sum\limits_{{\check x_1},{\check x_2}} {\frac{1}{{{{\left( {\sqrt {2\pi } \sigma } \right)}^{2r}}}}\prod\limits_{i = 1}^2 {\prod\limits_{t = 1}^r {\exp \left( { - \frac{{{{\left( {{z_{i,t}} + 2{\check x_{i,t}} + 2{y_{i,t}} - 2} \right)}^2}}}{{2{\sigma ^2}}}} \right)} } } .}
\end{array}
\end{equation}

Define ${{L'_s}} {\!}={\!} P\left( {z_1^2,{v_1}|u_1, \gamma^s} \right)$. Assuming $\tilde v_2 {\!}{\!}={\!}{\!} \gamma^s$ is transmitted, the probability of a correct decision is
\begin{equation}    \label{independent pc two user} 
{P_c} = {\!}{\!}{\!}{\!}\prod\limits_{{\gamma ^a} \in {\mathbb{F}_q},{\gamma ^a} \ne {\gamma ^s}} {\!}{\!}{\!}{P\left[ {{L'_s} > {L'_a}|{v_2} = {\gamma ^s}} \right].} 
\end{equation}

\vspace{-0.3em}   
It is hard to derive the term in (\ref{independent pc two user}) directly. Thus (\ref{sum hard probability}) is simplified by omitting the relatively small items. Assuming $\tilde u_2$ is transmitted, and $\tilde x_1, \tilde x_2$ correspond to $\tilde u_2$. Let $\bar v_2$ denote $\gamma^a$. Define $\tilde w_i = \tilde x_i + \tilde y_i$ and $\bar w_i = \check x_i + \bar y_i$, $1 \le i \le 2$, besides
\vspace{-0.5em}    
\begin{equation}
\bar x_1^2 = \mathop {\arg \min }\limits_{{\check x_1},{\check x_2} \in \mathbb{F}_q} \sum\limits_{i = 1}^2 {\left\| {{{\tilde w}_i} - {{\bar w}_i}} \right\|_2^2},
\end{equation}
where $\bar y_1, \bar y_2$ and $\tilde y_1, \tilde y_2$ correspond to $\bar v_2$ and $\tilde v_2$, respectively. Then (\ref{sum hard probability}) is approximated by
\begin{equation}  
\begin{array}{l}
P\left( {z_1^2,{v_1}|{u_1},{\bar v_2}} \right) \approx \\
\frac{1}{{{{\left( {\sqrt {2\pi } \sigma } \right)}^{2r}}}}\prod\limits_{i = 1}^2 {\prod\limits_{t = 1}^r {\exp \left( { - \frac{{{{\left( {{z_{i, t}} + 2{ \bar x_{i, t}} + 2{ y_{i, t}} - 2} \right)}^2}}}{{2{\sigma ^2}}}} \right)} }. 
\end{array}
\end{equation}

The situation degraded to the single-user case, and the average BER is calculated by going through all $u_2$ and $v_2$. Then the kernel factor $\alpha_v$ that provides the best BER is chosen.

\section{SIMULATION RESULTS}
\begin{figure}[t]
\centerline{\includegraphics[width=0.4\textwidth]{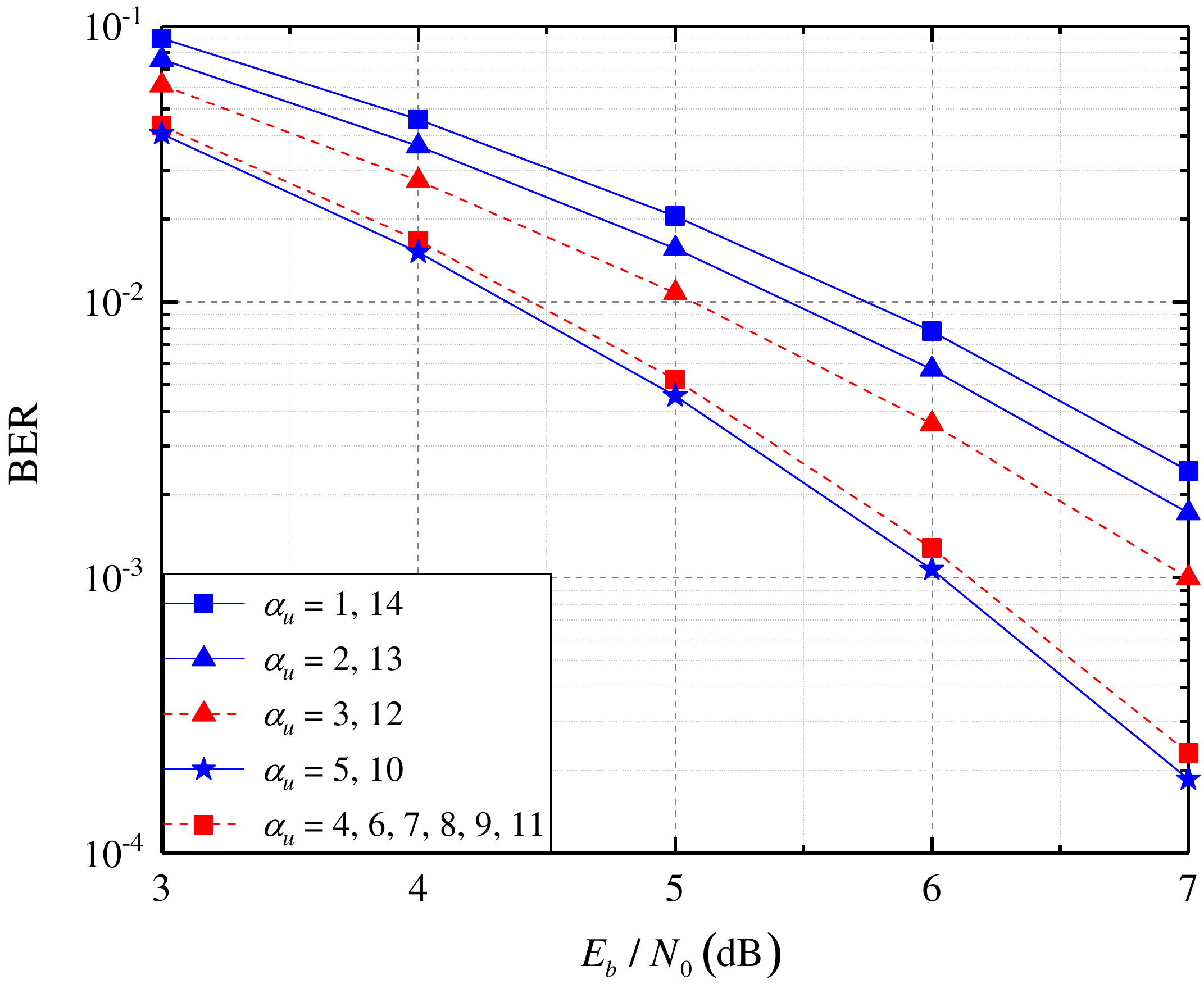}}
\vspace{-1em}   
\caption{Different kernels' impact on the single-user basic structure.}
\vspace{-0.6em}   
\label{different kernel single}
\end{figure}

\begin{figure}[t]
\centerline{\includegraphics[width=0.4\textwidth]{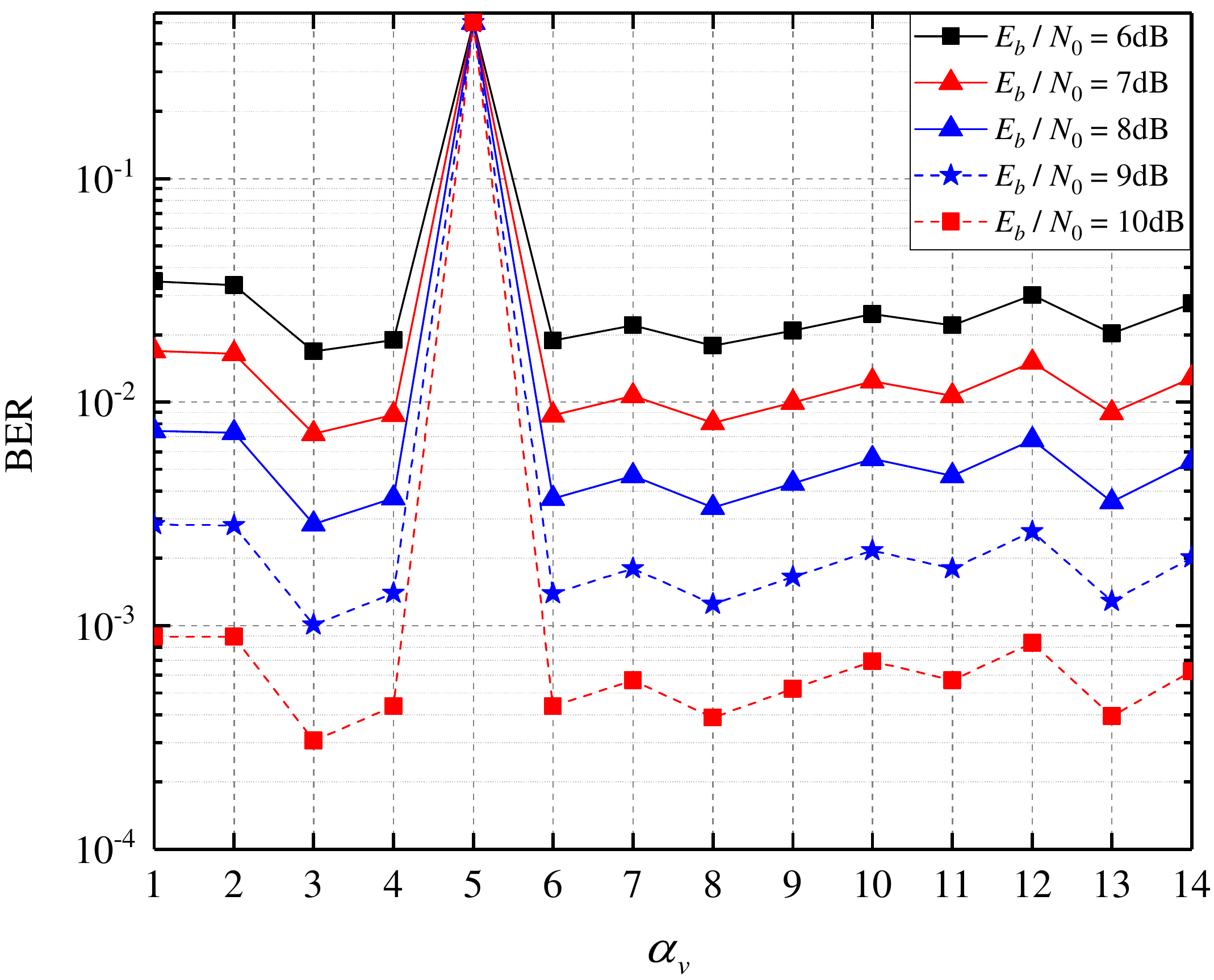}}
\vspace{-1em}   
\caption{Different kernels' impact on the combined basic structure.} 
\label{different kernel two user}
\end{figure}

In this section, Monte Carlo method is used to simulate the BLER performances of the proposed system. In this work, we take the ${\rm GF}\left({16}\right)$ as an example to do analysis, and the theoretical BERs of different kernels are shown in Figs. \ref{different kernel single} and \ref{different kernel two user}. According to Fig. \ref{different kernel single}, $\alpha_u = 5, 10$ is the optimal choice for the single-user case in ${\rm GF}\left({16}\right)$. Thus let $\alpha_u = 5$ be the kernel factor for user 1. According to Fig. \ref{different kernel two user}, $\alpha_v = 3$ is the optimal choice in the two-user case when $\alpha_u = 5$. Thus let $\alpha_v = 3$ be the kernel factor for user 2.

\begin{figure}[t]
\centerline{\includegraphics[width=0.4\textwidth]{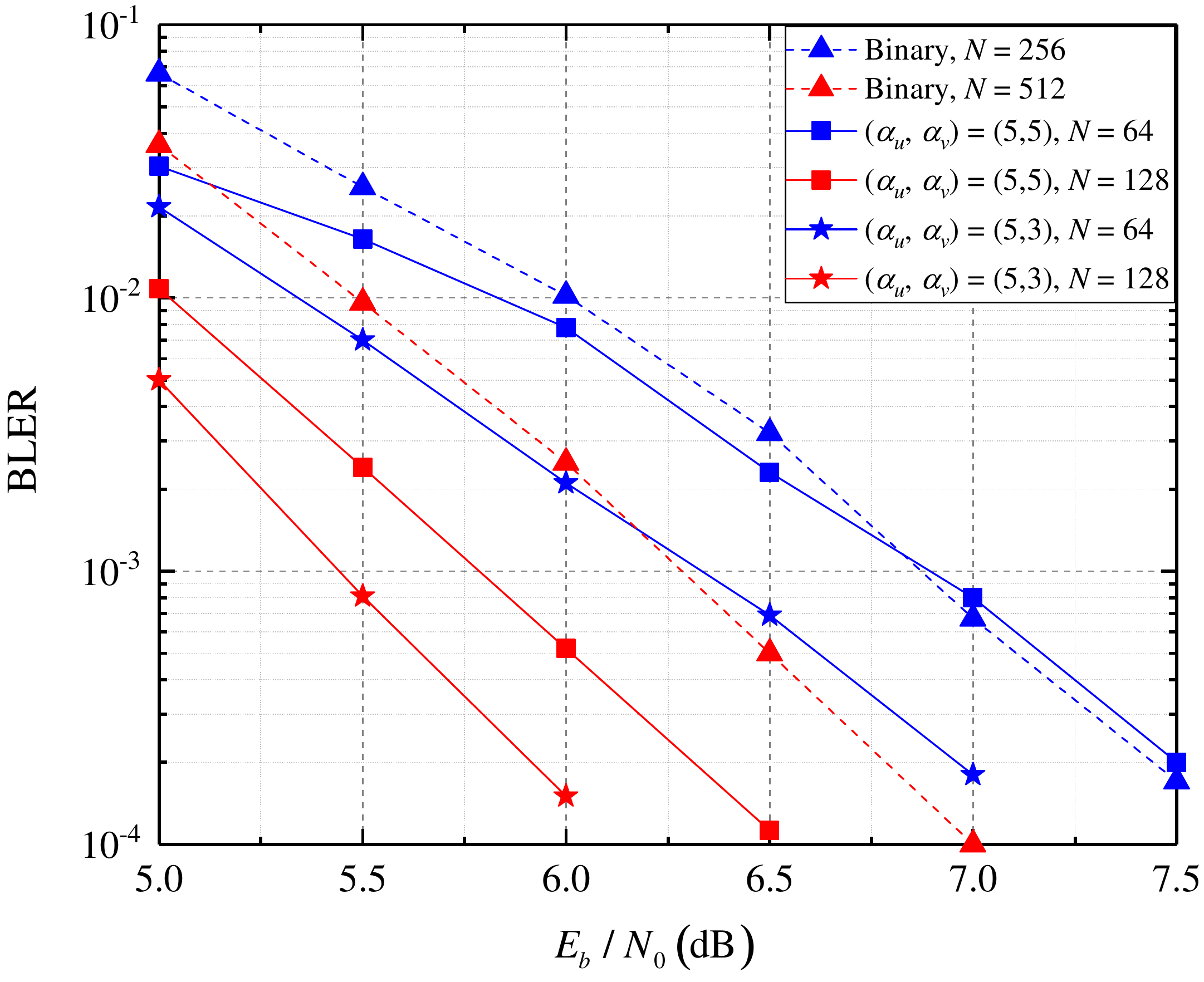}}
\vspace{-1em}   
\caption{BLER performance between non-binary and binary polar system.} 
\vspace{-1em} 
\label{BLER performance}
\end{figure}

Fig. \ref{BLER performance} compares the BLER performances between the proposed non-binary and binary systems in the two-user MAC with different code lengths, where the Monte Carlo construction is used to pick the transmission channels. The decoding order parameter $M$ is set as ${N \mathord{\left/
 {\vphantom {N 4}} \right.
 \kern-\nulldelimiterspace} 4}$, with $R=0.5$. First of all, the BLER performance comparations between two non-binary cases are considered, where $\left({\alpha_u, \alpha_v}\right)=\left({5, 3}\right)$ and $\left({\alpha_u, \alpha_v}\right)=\left({5, 5}\right)$, corresponding to the best and the worst case in ${\rm GF}\left({16}\right)$, given by Fig. \ref{different kernel two user}. When BLER $= 1 \times 10^{-3}$, the required $E_b/N_0$ of $N = 64, 128$ are respectively 6.3dB and 5.4dB for $\left({5, 3}\right)$ kernel case. Obviously, the BLER performance improves with the increase of $N$. Moreover, it is found that the BLER of $\left({5, 3}\right)$ case is better than that of $\left({5, 5}\right)$ case, e.g., there is 0.4dB gain when BLER $ = 1 \times 10^{-3}$ and $N = 128$, which is in line with the theoretical analysis. Secondly, it is found that the proposed $\left({5, 3}\right)$ non-binary polar system provides a much lower BLER performance than that of the binary polar system, e.g., when BLER $= 1 \times 10^{-3}$ and $N = 64, 128$, the $E_b/N_0$ of $\left({5, 3}\right)$ case has 0.5dB and 0.8dB gain, respectively. In summary, the non-binary polar system with the optimal kernel achieves relatively superior performance compared to the classical binary system.

\section{Conclusion}
This paper proposes a non-binary polar coding scheme in the two-user MAC and the corresponding SC-decoding algorithm. The choice of the kernel factors is discussed in detail. Simulation results show that there is a vast improvement between the worst and the best kernel choice. Moreover, the non-binary polar codes in the two-user MAC achieve much better BLER performances than the classical binary codes.

\vspace{12pt}

\end{document}